\documentclass[11pt]{article}
\usepackage{times}
\usepackage{mathptm}
\usepackage[colorlinks=true]{hyperref}
\newcommand{\enl}{\\\hfill\rule{0pt}{0pt}}
\begin{document}

\begin{center}
{{\Huge\bf BritGravII}\linebreak

{\Huge\bf Second British Gravity Meeting}\linebreak

{\large\bf School of Mathematical Sciences, Queen Mary, University
of London}\linebreak {\large\bf Mon/Tue, June 10/11,
2002}}

\end{center}

\noindent
The second of the annual {\bf BritGrav} meetings on current
research in Gravitational Physics in Britain took place at the
School of Mathematical Sciences of Queen Mary, University of London
on June 10/11, 2002. We tried to maintain the two good practices
established at Southampton last year (see Carsten Gundlach's report
at \href{http://arxiv.org/abs/gr-qc/0104062}{gr-qc/0104062});
namely allowing short talks of equal duration and keeping the
participants' expenses to a minimum. We were also able, with
support from the groups below, to provide partial financial support
for the younger participants.

Altogether 95 participants took part, including a good number from
research institutions outside Britain (namely, France, Sweden,
Russia, Spain, Austria, USA, Portugal, Ireland, Germany, South
Africa, Canada and Poland). There were a total of 47 talks --- of
12 minutes duration each --- over these two days. These included 1
talk by an undergraduate finalist, 15 talks by PhD students and 11
talks by postdoctoral researchers. The sessions were roughly
grouped into the following categories: Classical General
Relativity, Mathematical Studies, Quantum Gravitation, Quantum
Theory on Curved Spacetimes, Alternative Models, Relativistic
Astrophysics, Numerical Methods, and Other Topics.

Below is the list of the abstracts of all the talks, in
alphabetical order, submitted by the authors. The references to the
electronic preprints at \href{http://arxiv.org}{arXiv.org} have
been added where they exist.

The {\bf BritGravII} meeting was kindly supported financially by
the \href{http://www.lms.ac.uk}{London Mathematical Society}, the
\href{http://www.iop.org/IOP/Groups/MP/}{{\em Institute of
Physics\/} Mathematical and Theoretical Physics Group}, the
{\href{http://www.iop.org/IOP/Groups/GP/}{{\em Institute of
Physics\/} Gravitational Physics Group}}, the scientific journal
\href{http://www.iop.org/journals/cqg}{{\em Classical and Quantum
Gravity\/}}, and the \href{http://www.maths.qmul.ac.uk/}{School of
Mathematical Sciences, Queen Mary, University of London}.

The {\bf BritGravIII} meeting is to be organised by Robin Tucker of
the Department of Physics at the University of Lancaster and held
in September 2003. All enquiries should be directed to him
electronically at {\tt R.Tucker@lancaster.ac.uk}. \linebreak

\noindent
Henk van Elst and
Reza Tavakol, Astronomy Unit, Queen Mary, University of London

\vspace{3mm}
\begin{center}
{\huge\bf Submitted abstracts}
\end{center}

\begin{enumerate}

\item {\bf Anderson, Edward} \enl
(eda@maths.qmul.ac.uk, Astronomy Unit, Queen Mary, University of
London)

{\bf The 3-space approach to relativity: results\/} \enl
We show how 3 branches of gravitational theory arise from the
3-space approach: general relativity, arbitrary supermetric strong
gravity and a conformal theory of gravity [1,2,3].  We show then
how inclusion of vector fields in the general relativity branch
leads to electromagnetism and Yang-Mills theory [1,4].  The
technique used throughout is an exhaustive extension of Dirac's
generalized Hamiltonian formalism [4,5]. We finally comment on work
in progress contrasting our matter results in general relativity
with Kuchar's hypersurface formulation (which, in contrast to the
3-space approach, presupposes 4-geometry) [6].

    [1] J. Barbour, B.Z. Foster and N. O'Murchadha,
        \href{http://arxiv.org/abs/gr-qc/0012089}{gr-qc/0012089}
        (accepted by Class. Quantum Grav.),
    [2] J. Barbour and N. O'Murchadha,
        \href{http://arxiv.org/abs/gr-qc/9911071}{gr-qc/9911071},
    [3] E. Anderson,
        \href{http://arxiv.org/abs/gr-qc/0205118}{gr-qc/0205118},
    [4] E. Anderson and J. Barbour,
        \href{http://arxiv.org/abs/gr-qc/0201092}{gr-qc/0201092} 
        (accepted by Class. Quantum Grav.),
    [5] P.A.M. Dirac, Lectures in Quantum Mechanics 
        (Yeshiva University, NY 1964),
    [6] K. Kuchar, J.~Math.~Phys {\bf 17}, 777, 792, 801 (1976);
        J.~Math.~Phys. {\bf 18}, 1589 (1977).


\item {\bf Arminjon, Mayeul} \enl
(arminjon@hmg.inpg.fr, Laboratoire "Sols, Solides, Structures",
CNRS, Grenoble, France)

{\bf Asymptotic PN approximation and the point-particle limit in a
scalar theory of gravitation} \enl
To test celestial mechanics in an alternative theory based on just
a scalar field, an "asymptotic" PN scheme has been developed:
Applying the usual method of asymptotic expansions for a system of
PDE's, all fields are expanded in powers of the field-strength
parameter $\lambda$ - not merely the gravitational field as in the
standard PN scheme. This gives separate equations at each order in
$\lambda$. By integrating these local equations inside the bodies,
one gets equations of motion for the mass centers. The good
separation between bodies is exploited by introducing another small
parameter $\eta$ and truncating the equations of motion with
respect to $\eta$. The point particle limit is defined by assuming
that the size of all bodies but one shrinks with a small parameter
$\xi$. The equation of motion obtained thus contains a
structure-dependent term and does not coincide with the equation of
motion for a test particle in the field of the massive body. This
follows from using the asymptotic scheme and presumably holds true
for GR.

{\bf arXiv.org\/}:
\href{http://arxiv.org/abs/gr-qc/0202029}{gr-qc/0202029}

\item {\bf Barbour, Julian} \enl
(julian@platonia.com, Banbury)

{\bf The 3-space approach to relativity: motivation} \enl
In a recent paper, 'Relativity without relativity', Barbour, Foster
and O'Murchadha have presented a new approach to relativity and
gauge theory based entirely on three-dimensional concepts. My talk
will motivate this 3-space approach and show how it has the
potential to clarify difficult conceptual issues in quantum
gravity, ease some of the technical difficulties, and perhaps lead
to a viable scale-invariant generalization of general relativity.

{\bf arXiv.org\/}:
\href{http://arxiv.org/abs/gr-qc/0012089}{gr-qc/0012089},
\href{http://arxiv.org/abs/gr-qc/9911071}{gr-qc/9911071}

\item {\bf Bonnor, William} \enl
(100571.2247@compuserve.com, School of Mathematical Sciences, Queen
Mary, University of London)

{\bf Closed timelike curves} \enl
Closed timelike curves (CTC) occur in general relativity, raising
the question of time travel and its associated paradoxes.  They are
usually dismissed on the grounds that the spacetimes in which they
arise are non-physical.  However, this is not so in all cases.

I argue that relativists ought seriously to consider the
significance of CTC.


\item {\bf Carr, Bernard} \enl
(B.J.Carr@qmul.ac.uk, Astronomy Unit, Queen Mary, University of
London)

{\bf Tolman--Bondi collapse in scalar-tensor theory as a probe of
gravitational memory} \enl
In cosmological models with a varying gravitational constant, it is
not clear whether primordial black holes preserve the value of $G$
at their formation epoch. We investigate this question by using the
Tolman--Bondi model to study the evolution of a background scalar
field when a black hole forms from the collapse of dust in a flat
Friedmann Universe. Providing the back reaction of the scalar field
on the metric can be neglected, we find that the value of the
scalar field at the event horizon very quickly assumes the
background cosmological value. This suggests that there is very
little gravitational memory.

{\bf arXiv.org\/}:
\href{http://arxiv.org/abs/astro-ph/0112563}{astro-ph/0112563}

\item {\bf Docherty, Peter} \enl
(Thepdochertys@btopenworld.com, Department of Mathematical
Sciences, University of Loughborough)

{\bf A disintegrating cosmic string} \enl
We present a simple sandwich gravitational wave of the
Robinson--Trautman family. This is interpreted as representing a
shock wave with a spherical wavefront which propagates into a
Minkowski background minus a wedge. (i.e. the background contains a
cosmic string.) The deficit angle (the tension) of the string
decreases through the gravitational wave, which then ceases. This
leaves an expanding spherical region of Minkowski space behind
it. The decay of the cosmic string over a finite interval of
retarded time may be considered to generate the gravitational wave.

{\bf arXiv.org}:
\href{http://arxiv.org/abs/gr-qc/0204085}{gr-qc/0204085}

\item {\bf Dolby, Carl} \enl
(dolby@ariadne.physics.ox.ac.uk, Theoretical Physics, University of
Oxford)

{\bf Radar time, simultaneity, and QFT for abitrary observers in
gravitational backgrounds} \enl
An approach to fermionic QFT in gravitational and electromagnetic
backgrounds will be presented, which allows a consistent particle
interpretation at all `times'. The concept of radar time
(popularised by Bondi in his work on k-calculus) provides a
suitable observer dependent foliation of spacetime, hence allowing
this interpretation to be applied to an arbitrarily moving
observer. A number operator results, which depends only on the
observers motion and the background present (not on any choice of
coordinates or gauge), and which generalises Gibbons' definition to
non-stationary spacetimes. This operator is non-local on small
scales, but is `effectively local' on scales larger than the
Compton wavelength of the particle concerned. Applications of this
construction to spatially uniform electric fields, accelerating
observers, and simple cosmologies will be presented.

{\bf arXiv.org}:
\href{http://arxiv.org/abs/hep-th/0103228}{hep-th/0103228},
\href{http://arxiv.org/abs/gr-qc/0104077}{gr-qc/0104077}

\item {\bf Dowker, Fay} \enl
(f.dowker@qmul.ac.uk, Department of Physics, Queen Mary, University
of London)

{\bf The deep structure of spacetime} \enl
A brief review of the causal set approach to quantum gravity.


\item {\bf Edgar, Brian} \enl
(bredg@mai.liu.se, Mathematics Department, University of
Link\"{o}ping, Sweden)

{\bf Applications of dimensionally dependent identities} \enl
Dimensionally dependent identities are tensor identities which are
valid only in some dimension(s).  As well as providing a unifying
derivation of familiar, but apparently unrelated identities, they
are important tools for a systematic treatment --- in different
dimensions --- of such topics as invariants of the Riemann tensor,
and properties of super-energy tensors.

{\bf arXiv.org}:
\href{http://arxiv.org/abs/gr-qc/0105066}{gr-qc/0105066},
\href{http://arxiv.org/abs/gr-qc/0202092}{gr-qc/0202092}

\item {\bf Fil'chenkov, Michael} \enl
(fil@crosna.net, Institute of Gravitation and Cosmology, Peoples'
Friendship University of Russia, Moscow, Russia)

{\bf Quantum birth of a universe with rotation} \enl
Of importance is the question of rotation of the early Universe
despite the observational angular velocity is now negligible, if
exists. Integrating Raychaudhuri's equation for a homogeneous
multicomponent universe with rotation, we obtain a Friedmann-like
equation. Quantizing the latter gives Wheeler--DeWitt's
equation. Energy levels of the pre-de-Sitter universe as well as a
tunnelling factor for the birth of a rotating universe are
calculated. Black-body radiation and de Sitter's vacuum do not take
part in the rotation of the Universe. The tunnelling factor proves
to depend on the angular momenta of matter components. The general
problem of the origin of rotation in astronomy is discussed in
connection with its possible cosmological nature as well as the
existing formulae for angular momenta of astronomical objects. The
quantum primordial rotation may affect a further classical
evolution of the Universe, generating the observable angular
momenta of galaxies.


\item {\bf Gaburov, Evghenii} \enl
(eg35@leicester.ac.uk, Department of Physics and Astronomy,
University of Leicester)

{\bf Anisotropic black holes in Einstein and brane gravity} \enl We
consider exact solutions of Einstein equations defining static
black holes parametrized by off-diagonal metrics which by
anholonomic mappings can be equivalently transformed into some
diagonal metrics with coefficients being very similar to those from
the Schwarzschild and/or Reissner--N\"ordstrom solutions with
anisotropic renormalizations of constants. We emphasize that such
classes of solutions, for instance, with ellipsoidal symmetry of
horizons, can be constructed even in general relativity theory if
off-diagonal metrics and anholonomic frames are introduced into
considerations. Such solutions do not violate the Israel's
uniqueness theorems on static black hole configurations because at
long radial distances one holds the usual Schwarzschild limit. We
show that anisotropic deformations of the Reissner--N\"ordstrom
metric can be an exact solution on the brane, re-interpreted as a
black hole with an effective electromagnetic like charge
anisotropically induced and polarized by higher dimension
gravitational interactions.

{\bf arXiv.org}:
\href{http://arxiv.org/abs/hep-th/0108065}{hep-th/0108065}

\item {\bf Garcia-Islas, Juan Manuel} \enl
(jm.garcia-islas@maths.nottingham.ac.uk, School of Mathematical
Sciences, University of Nottingham)

{\bf Observables in 2+1 dimensional Euclidian quantum gravity} \enl
A new set of observables is defined in the framework of 2+1
dimensional quantum gravity. This is studied using the Turaev--Viro
model which is the deformed version of the Ponzano--Regge model, as
it uses the powerful methods of quantum groups. The expectation
values of our new observables turn out to be related with
relativistic spin network invariants. Moreover, when considering
some particular examples it appears that we can obtain some
relationships with a rational conformal field theory in the form of
identities.


\item {\bf Garc\'{\i}a-Parrado G\'{o}mez-Lobo, Alfonso} \enl
(wtbgagoa@lg.ehu.es, Departamento de F\'{\i}sica Te\'{o}rica e
Historia de la Ciencia, Universidad del Pa\'{\i}s Vasco, Bilbao,
Spain)

{\bf Causal relationship: a new tool for the causal
characterization of Lorentzian manifolds} \enl
We define a new kind of relation between two diffeomorphic
Lorentzian manifolds called {\em causal relation}, which is any
diffeomorphism characterized by mapping every future vector of the
first manifold onto a future vector of the second.  We perform a
thorough study of the mathematical properties of causal relations
and prove in particular that two given Lorentzian manifolds may be
causally related only in one direction.  This leads us to the
concept of causally equivalent Lorentzian manifolds as those
mutually causally related. This concept is more general and of a
more basic nature than the conformal relationship, because we prove
the remarkable result that a conformal relation $f$ is
characterized by the fact of being a causal relation of the {\em
particular} kind in which both $f$ and $f^{-1}$ are causal
relations.  Another important feature of causally equivalent
Lorentzian manifolds is that there is a one-to-one correspondence
between their respective future (and past) objects. This clearly
indicates that the causal equivalence is the right definition
capturing the {\it causal indistinguishability} of Lorentzian
manifolds. Further, it is possible to introduce a partial order in
the set of Lorentzian manifolds providing a classification of
classes of spacetimes in terms of their causal properties.

A parallel study for embedded hypersurfaces in Lorentzian manifolds
is also carried out, which together with the concept of {\em
asymptotically isocausal} spacetimes, allows us to attach a causal
boundary to some spacetimes.  Explicit examples will be presented.

{\bf arXiv.org}:
\href{http://arxiv.org/abs/math-ph/0202005}{math-ph/0202005}

\item {\bf Griffiths, Jerry} \enl
(J.B.Griffiths@lboro.ac.uk, Department of Mathematical Sciences,
University of Loughborough)

{\bf The initial value problem for colliding plane waves: the
nonlinear case} \enl
A general method is presented for constructing solutions for
colliding plane pure gravitational or mixed gravitational and
electromagnetic waves with distinct wavefronts which propagate
initially into a Minkowski background. This method enables us in
principle to construct the solution in the wave interaction region
directly in terms of initial data that are specified on two null
characteristics. This is achieved by reducing the characteristic
initial value problem to some simple linear integral ``evolution''
equations. The method presented for solving these equations arises
from a general monodromy transform approach. This makes use of the
fact that the symmetry reduced Einstein or Einstein--Maxwell
equations (which are equivalent to the hyperbolic Ernst equations)
can be reformulated as linear integral equations. These can be
constructed in terms of ``dynamical'' monodromy data for solutions
of an associated linear problem on the spectral plane. It is shown
that such an approach can be generalized to solutions with
nonanalytical behaviour of fields on null characteristics as is
required for colliding waves with distinct wavefronts. Using a
number of examples, it is demonstrated how the method presented can
be used in practice to solve this nonlinear characteristic initial
value problem for colliding plane waves.


\item {\bf Grumiller, Daniel} \enl
(grumil@hep.itp.tuwien.ac.at, Institute for Theoretical Physics,
Vienna University of Technology, Austria)

{\bf Virtual black hole phenomenology from 2d dilaton theories}
\enl
Equipped with the tools of (spherically reduced) dilaton gravity in
first order formulation and with the results for the lowest order
S-matrix for s-wave gravitational scattering, new properties of the
ensuing cross-section are discussed. We find CPT invariance,
despite of the non-local nature of our effective theory and
discover pseudo-self-similarity in its kinematic sector. After
presenting the Carter--Penrose diagram for the corresponding virtual
black hole geometry we encounter distributional contributions to
its Ricci-scalar and a vanishing Einstein--Hilbert action for that
configuration. Finally, a comparison is done between our
(Minkowskian) virtual black hole and Hawking's (Euclidean) virtual
black hole bubbles.

{\bf arXiv.org}:
\href{http://arxiv.org/abs/gr-qc/0111097}{gr-qc/0111097},
\href{http://arxiv.org/abs/gr-qc/0105034}{gr-qc/0105034}

\item {\bf Hall, Graham} \enl
(g.hall@maths.abdn.ac.uk, Department of Mathematical Sciences,
University of Aberdeen)

{\bf Sectional curvature in general relativity} \enl
Let $(M,g)$ be a space-time, let $p \in M$ and let $T_pM$ be the
tangent space to $M$ at $p$. Let $F$ be a non-null $2$-space at $p$
spanned by $u,v \in T_pM$. Define, in a standard notation, the
\emph{sectional curvature} of $F$ by
\[ \sigma_p(F) =
\frac{R_{abcd}F^{ab}F^{cd}}{2g_{a[c}g_{d]b}F^{ab}F^{cd}} \qquad
(F_{ab} = 2u_{[a}v_{b]}). \]
This definition depends only on $F$ and not on the spanning pair
$u$ and $v$. Now define a real valued function $\sigma$ on the set
of all non-null $2$-spaces at all points of $M$ according to
$\sigma(F) = \sigma_p(F)$ if $F$ is a non-null $2$-space at $p$.

A detailed study of the function $\sigma$ can be carried out and
reveals that given that $\sigma_p$ is not a constant function for
any $p$ and that $(M,g)$ is not a conformally flat plane-wave, then
the function $\sigma$ uniquely determines the metric $g$ from
whence it came. In particular, for nowhere flat vacuum space-times,
the metric $g$ and the sectional curvature function $\sigma$ are in
one-to-one correspondence. Thus the sectional curvature function
could be used as an alternative field variable for general
relativity, at least in the vacuum situation.

The sectional curvature function can also be used to describe
symmetries on $M$, and the Petrov type and energy momentum tensor
type at each point of $M$ can be recovered from it.


\item {\bf Henson, Joe} \enl
(j.j.henson@qmul.ac.uk, Department of Physics, Queen Mary,
University of London)

{\bf What are the ``observables'' of quantum gravity?} \enl
A brief discussion of what the different approaches to quantum
gravity hope will be the most general sorts of questions they can
answer.  An explanation of why the problem has a well defined
meaning and possible solution in the casual set approach.


\item {\bf Jones, David Ian} \enl
(D.I.Jones@maths.soton.ac.uk, Faculty of Mathematical Studies,
University of Southampton)

{\bf Numerical implementation of a local radiation reaction} \enl
In this talk I will present the results of a linear Newtonian
hydrodynamics code in which the dissipative effect of gravitational
wave emission is modelled using a local force.  The force itself is
that derived via post-Newtonian methods by Blanchet, Damour and
Schafer, who eliminated the high time derivatives found in other
formulations by introducing a number of Poisson-like equations.  It
is hoped that by testing this method in the linear regime it can
then be easily extended and used with confidence is the non-linear
one.


\item {\bf Konkowski, Deborah} \enl
(dak@usna.edu, US Naval Academy, Annapolis, Maryland, USA)

{\bf Definition and classification of singularities in GR:
classical and quantum} \enl
After briefly reviewing the definition and classification of
classical singularities in general relativistic spacetimes, I will
give a definition of a quantum singularity in a spacetime following
the pioneering work of G. Horowitz and D. Marolf. Examples of
classically singular spacetimes which do and do not have quantum
singularities will be given. If time permits I will present results
on quasiregular spacetimes [Konkowski, D.A. and Helliwell,
T.M. (2001), Gen. Rel. Grav. 33, 1131] and static
cylindrically-symmetric spacetimes [work in progress].


\item {\bf Lambert, Paul} \enl
(p.e.lambert@maths.soton.ac.uk, Faculty of Mathematical Studies,
University of Southampton)

{\bf The constraint algebra of the $2+2$ Ashtekar Hamiltonian} \enl
In this talk I present the results of applying the Dirac--Bergman
algorithm to a $2+2$ (double null) Lagrangian based on self-dual
2-forms described in [1]. The use of Ashtekar variables results in
polynomial constraints. The first class constraint algebra forms a
Lie algebra. I will then give some discussion of the geometrical
meaning of the first class constraints. This work provides the
first stage of a canonical quantisation procedure for the Einstein
equations.

[1] R.A. d'Inverno and J.A. Vickers \emph{2+2
decomposition of Ashtekar variables} Class. Quantum Grav. {\bf 12},
753 (1995).


\item {\bf Lazkoz Saez, Ruth} \enl
(wtplasar@lg.ehu.es, Departamento de F\'{\i}sica Te\'{o}rica e
Historia de la Ciencia, Universidad del Pa\'{\i}s Vasco, Bilbao,
Spain)

{\bf Newtonian limit of boost-rotation symmetric spacetimes} \enl
Boost-rotation spacetimes are believed to describe particles in
hyperbolic motion. We study the Newtonian limit of those spacetimes
within the context of Cartan--Friedrich frame theory and provide
general results concerning conditions of the existence of the limit
of the corresponding time and space-metrics, connection are Riemann
tensor. This leads to a clear identification of the Newtonian
potential of those boost-rotation spacetimes admitting such limits
and in particular of some examples.


\item {\bf Louko, Jorma} \enl
(Jorma.Louko@nottingham.ac.uk, School of Mathematical Sciences,
University of Nottingham)

{\bf Thermal effects with a locally bifurcate Killing horizon} \enl
Thermal effects in quantum field theory on spacetimes with a
bifurcate Killing horizon, such as nonextremal black holes or
Rindler space, are usually attributed to the existence of two
causally disconnected `exterior' regions separated by the
horizon. We discuss the situation in spacetimes that have a local
analogue of a bifurcate Killing horizon and just one `exterior'
region. Thermal effects are then no longer exact but emerge
approximately in certain limits, including the limits of late and
early times.

{\bf arXiv.org}:
\href{http://arxiv.org/abs/hep-th/0002111}{hep-th/0002111},
\href{http://arxiv.org/abs/gr-qc/9906031}{gr-qc/9906031},
\href{http://arxiv.org/abs/gr-qc/9812056}{gr-qc/9812056},
\href{http://arxiv.org/abs/hep-th/9808081}{hep-th/9808081},
\href{http://arxiv.org/abs/gr-qc/9802068}{gr-qc/9802068}

\item {\bf Martin, Nigel} \enl
(gnnmartin@cix.compulink.co.uk)

{\bf The unambiguous speed of light} \enl
This paper observes that the field equations of relativity, in the
form $T=G$, represent a reasonable generalisation of the definition
of mass in the light of an assumption that the local speed of light
is constant.  The hope is that this allows relativity to be
explained without compromise to a less mathematical audience,
building up logically from ideas that will be familiar to any
ordinary well educated person.


\item {\bf Martin-Garcia, Jos\'{e}} \enl
(J.M.Martin-Garcia@maths.soton.ac.uk, Faculty of Mathematical
Studies, University of Southampton)

{\bf Self-similarity in the Vlasov--Einstein system} \enl
The Vlasov--Einstein system describes the evolution of a
statistical ensemble of non-interacting particles coupled to
gravity through their average properties. It is the only system
where `critical phenomena in gravitational collapse' have not been
found. In this talk we address this problem, restricting the study
to the simpler case of massless particles. Only one of the two key
ingredients for criticality is present: there are self-similar
solutions, but they cannot be codimension-1 stable.

{\bf arXiv.org}:
\href{http://arxiv.org/abs/gr-qc/0112009}{gr-qc/0112009}

\item {\bf Matteucci, Paolo} \enl
(p.matteucci@maths.soton.ac.uk, Faculty of Mathematical Studies,
University of Southampton)

{\bf Multisymplectic derivation of $2+2$ Hamiltonian dynamics} \enl
Multisymplectic geometry, which stems from some pioneering work of
De Donder and Weyl in the early 1930s, has recently come back into
vogue owing to an innovative approach to quantization proposed by
Kanatchikov. Unlike the ADM approach, where dynamics is described
in terms of the infinite-dimensional space of fields at a given
instant of time, in the multisymplectic formalism dynamics is
phrased in the context of the finite-dimensional space of fields at
a given event in space-time. The 2+2 formalism developed by
d'Inverno, Stachel and Smallwood, which is particularly suited to
many situations in General Relativity, lies exactly in between the
two approaches, and, whereas its relationship with the former is
well-known, there is reason to believe that only a correct
understanding of its connection with the latter will provide full
insight into its geometry.


\item {\bf Mena, Filipe} \enl
(fmena@math.uminho.pt, Departamento de Matematica, Universidade do
Minho, Braga, Portugal)

{\bf Initial data and spherical dust collapse} \enl
We study the role of the initial data in the final state of
collapse in Lema\^{\i}tre--Tolman--Bondi models. In connection to
the cosmic censorship conjecture we study the existence of null
radial geodesics which emanate from the central singularity. We
consider stability aspects of the black hole and naked singularity
solutions.

{\bf arXiv.org}:
\href{http://arxiv.org/abs/gr-qc/0002062}{gr-qc/0002062},
\href{http://arxiv.org/abs/gr-qc/0108008}{gr-qc/0108008}

\item {\bf Mukohyama, Shinji} \enl
(mukoyama@schwinger.harvard.edu, Department of Physics, Harvard
University, Cambridge (MA), USA)

{\bf Brane cosmology driven by the rolling tachyon} \enl
Brane cosmology driven by the tachyon rolling down to its ground
state is investigated. We adopt an effective field theoretical
description for the tachyon and Randall--Sundrum type brane world
scenario. After formulating basic equations, we show that the
standard cosmology with a usual scalar field can mimic the low
energy behavior of the system near the tachyon ground state. We
also investigate qualitative behavior of the system beyond the low
energy regime for positive, negative and vanishing 4-dimensional
effective cosmological constant
$\Lambda_4=\kappa_5^4V(T_0)^2/12-|\Lambda_5|/2$, where $\kappa_5$
and $\Lambda_5$ are 5-dimensional gravitational coupling constant
and (negative) cosmological constant, respectively, and $V(T_0)$ is
the (positive) tension of the brane in the tachyon ground state. In
particular, for $\Lambda_4<0$ the tachyon never settles down to its
potential minimum and the universe eventually hits a big-crunch
singularity.

{\bf arXiv.org}:
\href{http://arxiv.org/abs/hep-th/0204084}{hep-th/0204084}

\item {\bf Nerozzi, Andrea} \enl
(andrea.nerozzi@port.ac.uk, Institute of Cosmology and Gravitation,
University of Portsmouth)

{\bf Relativistic irrotational fluids: 3D simulations in
Schwarzschild metric} \enl
Irrotational fluids constitute an interesting topic in astrophysics
since the Euler equations reduce in this case to a non-linear
scalar field equation. Exact solutions for irrotational fluids
accreting onto black-holes have been found in literature in the
further approximation of the fluid sound speed being equal to the
speed of light, the equation being in this case linear.  I present
in my talk some 3D numerical simulations for matter accreting onto
a Schwarzschild black hole, in the non-linear regime in which no
constraint is imposed on the fluid sound speed.


\item {\bf Nolan, Brien} \enl
(Brien.Nolan@dcu.ie, Dublin City University, Ireland)

{\bf Generalized solutions for shell-crossing singularities} \enl
We derive generalized solutions of Einstein's equations for
spherically symmetric dust-filled space-times which admit
shell-crossing singularities. In the marginally bound case, the
solutions are weak solutions of a conservation law which is
equivalent to the field equation in this case. In the
non-marginally bound case, the equations are solved in a
generalized sense involving metric functions of bounded variation.
The solutions are not unique to the future of the shell-crossing
singularity, which is a shock wave in the present treatment; the
metric is bounded but not continuous.


\item {\bf Patel, Mohammed} \enl
(mpatel@maths.abdn.ac.uk, Department of Mathematical Sciences,
University of Aberdeen)

{\bf Projective symmetry in conformally flat perfect fluid
spacetimes} \enl
Following a study of projective symmetry in FRW models, it is shown
that the only conformally flat, perfect fluid spacetimes admitting
such symmetry are essentially (locally) of the FRW type.


\item {\bf Pfenning, Michael} \enl
(mjp11@york.ac.uk, Department of Mathematics, University of York)

{\bf Quantum inequalities for the electromagnetic and Proca fields
in hyperbolic spacetimes} \enl
It has been known for some time that all of the classical energy
conditions in general relativity can be violate by quantum
fields. Thus, the energy density can become arbitrarily
negative. This can lead to unobserved phenomena such as violations
of the second law of thermodynamics, repulsive gravity and the
creation of spacetimes with closed timelike curves. Fortunately,
quantum field theory also places strict limits on negative
energies. Where as the pointwise energy density may become
arbitrarily negative, weighted time averages, known as quantum
inequalities, are bounded below.  I will present recent results
which prove that such inequalities exist for spin one fields in
globally hyperbolic spacetimes.


\item {\bf Polnarev, Alexander} \enl
(A.G.Polnarev@qmul.ac.uk, Astronomy Unit, Queen Mary, University of
London)

{\bf Response of a spaceborn gravitational antenna to solar
oscillations} \enl
The possibility of observing very small amplitude low frequency
solar oscillations with the proposed laser interferometer space
antenna LISA is investigated. For frequencies below 0.0002 Hz the
dominant contribution is from the near zone time dependent
gravitational quadruple moments associated with the normal modes of
oscillation. For frequencies above 0.0003 Hz the dominant
contribution is from gravitational radiation generated by the
quadrupole oscillations. The low order solar quadrupole pressure
and gravity oscillation modes have not yet been detected above the
solar background by helioseismic velocity and intensity
measurements. The estimates of the amplitudes needed to give a
detectable signal on a LISA type space laser interferometer imply
surface velocity amplitudes on the sun of the order of 1-10 mm/sec
in the frequency range 0.0001-0.0005 Hz. Such surface velocities
are below the current sensitivity limits on helioseismic
measurements. If modes exist with frequencies and amplitudes in
this range they could be detected with a LISA type laser
interferometer.

{\bf arXiv.org}:
\href{http://arxiv.org/abs/astro-ph/0103472}{astro-ph/0103472}

\item {\bf Prix, Reinhard} \enl
(R.Prix@maths.soton.ac.uk, Faculty of Mathematical Studies,
University of Southampton)

{\bf Adiabatic oscillations of non-rotating superfluid neutron
stars} \enl
I present results concerning the linear adiabatic oscillations of
non--rotating superfluid neutron stars in Newtonian gravitation. A
two--fluid model is used to describe the superfluid neutron star,
where one fluid consists of the superfluid neutrons, while the
second fluid contains all the comoving constituents (protons,
electrons).  I show numerical results which indicate the doubling
of all "acoustic" modes (f- and p- modes), and confirm the absence
of g--modes in these superfluid models.  The properties of these
two--fluid modes change as functions of the coupling by
entrainment, and one generally finds avoided mode-crossings. The
oscillations of normal-fluid neutron stars are recovered as a
special case simply by locking the two fluids together.  In this
effective one-fluid case we find the usual singlet f- and p- modes,
and we also recover the expected g-modes of stratified neutron star
models.  The presence or absence of g-modes could therefore give a
direct observational indication of superfluidity in neutron stars.

{\bf arXiv.org}:
\href{http://arxiv.org/abs/astro-ph/0204520}{astro-ph/0204520}

\item {\bf Re, Virginia} \enl
(Virginia.Re@port.ac.uk, Institute of Cosmology and Gravitation,
University of Portsmouth)

{\bf How to invariantly characterize non-linear black hole
perturbations}  \enl
The aim of this work is to introduce a characterization of a non
linearly perturbed black-hole spacetime in vacuum, using an
approach based on the Weyl curvature scalars. The physical
background is the Bondi--Sachs metric.  This metric describes an
axisymmetric non-rotating spacetime and, from a purely physical
viewpoint, it can be considered as a ''perturbation'' of a
spherical black-hole described by the Schwarzschild metric.  Our
purpose is to characterize this non linear perturbations using the
Weyl scalars.  The Bondi metric is Petrov type I, that is the most
general case in which no particular physical characteristic
emerges, unless it is possible to fall in one of the so-called
standard forms for Petrov type I. From this point of view, the
physically interesting case is the one in which the scalars
$\Psi_1$ and $\Psi_3$ are equal to zero, because they represent
only gauge fields.  In order to obtain this, we use the three
classes of rotations for the tetrad vectors. Once obtained the
rotated non zero scalars $\Psi_0$, $\Psi_2$, $\Psi_4$, the aim is
to explicitly calculate them at each point of the spacetime using
the Bondi code in order to get an invariant characterization of the
curvature throughout the spacetime.


\item {\bf Roberts, Mark} \enl
(mdrobertsza@yahoo.co.uk, Wonersh Park)

{\bf The rotation and shear of a string} \enl
Whether a string has rotation and shear can be investigated by an
analogy with the point particle. Rotation and shear involve first
covariant spacetime derivatives of a vector field and, because the
metric stress tensor for both the point particle and the string
have no such derivatives, the best vector fields can be identified
by requiring the conservation of the metric stress. It is found
that the best vector field is a non-unit accelerating field in $x$,
rather than a unit non-accelerating vector involving the momenta;
it is also found that there is an equation obeyed by the spacetime
derivative of the Lagrangian. The relationship between membranes
and fluids is looked at.

{\bf arXiv.org}:
\href{http://arxiv.org/abs/hep-th/0204236}{hep-th/0204236}

\item {\bf Santano-Roco, Miguel} \enl
(M.Santano-Roco@lboro.ac.uk, Department of Mathematical Sciences,
University of Loughborough)

{\bf The characteristic initial value problem for colliding
plane waves: The linear case} \enl
The physical situation of the collision and subsequent interaction
of plane gravitational waves in a Minkowski background is a
well-posed characteristic initial value problem in which the
initial data is specified on the two null characteristics that
define the wavefronts. In this talk, there will be analysed how
the Abel transform method can be used in practice to solve this
problem for the linear case in which the polarization of the two
gravitational waves is constant and aligned. There will also be
shown how the method works for some known solutions, where the
problems arise in other cases, and how the problem can always be
solved in terms of an infinite series if two initial spectral
functions can be determined.

{\bf arXiv.org}:
\href{http://arxiv.org/abs/gr-qc/0206075}{gr-qc/0206075}

\item {\bf Singh, Dinesh} \enl
(d.singh@lancaster.ac.uk, Department of Physics, University of
Lancaster)

{\bf Scattering of spinning test particles by plane
gravitational and electromagnetic waves} \enl
The Mathisson--Papapetrou--Dixon (MPD) equations for the motion of
electrically neutral massive spinning particles are analysed, in
the pole-dipole approximation, in an Einstein--Maxwell plane-wave
background spacetime. By exploiting the high symmetry of such
spacetimes these equations are reduced to a system of tractable
ordinary differential equations. Classes of exact solutions are
given, corresponding to particular initial conditions for the
directions of the particle spin relative to the direction of the
propagating background fields. For Einstein--Maxwell pulses a
scattering cross section is defined that reduces in certain limits
to those associated with the scattering of scalar and Dirac
particles based on classical and quantum field theoretic
techniques. The relative simplicity of the MPD approach and its use
of macroscopic spin distributions suggests that it may have
advantages in those astrophysical situations that involve strong
classical gravitational and electromagnetic environments.

{\bf arXiv.org}:
\href{http://arxiv.org/abs/gr-qc/0203038}{gr-qc/0203038}

\item {\bf Sopuerta, Carlos F} \enl
(carlos.sopuerta@port.ac.uk, Institute of Cosmology and
Gravitation, University of Portsmouth)

{\bf Two-parameters non-linear spacetime perturbations} \enl
An underlying fundamental assumption in relativistic perturbation
theory is that there exists a parametric family of spacetimes that
can be Taylor expanded around a background.  The choice of the
latter is crucial to obtain a manageable theory, so that it is
sometime convenient to construct a perturbative formalism based on
two (or more) parameters.  The study of perturbations of rotating
stars is a good example: in this case it may be convenient to treat
the axisymmetric star using a slow rotation approximation
(expansion in the angular velocity), so that the background is
spherical.  We analyse the gauge dependence of non-linear
perturbations depending on two parameters, derive explicit higher
order gauge transformation rules, and define gauge invariance in
this context.


\item {\bf Steele, Christopher} \enl
(Christopher.Steele@maths.nottingham.ac.uk, School of Mathematical
Sciences, University of Nottingham)

{\bf Relativistic spin networks} \enl
I will discuss Relativistic Spin Networks based on the
representation theory of the four dimensional rotation group. I
will present recent work on the asymptotics of the Riemannian
4-simplex, extending work of Barrett and Williams [1]. Some
numerical results will be explained using stationary phase
calculations.

    [1] The asymptotics of an amplitude for the 4-simplex John
    W. Barrett, Ruth M. Williams
    \href{http://arxiv.org/abs/gr-qc/9809032}{gr-qc/9809032}.


\item {\bf Tucker, Robin} \enl
(r.tucker@lancaster.ac.uk, Department of Physics, University of
Lancaster)

{\bf The LASSO project} \enl
It is proposed to explore the interaction of weak gravitational
fields with slender elastic materials in order to assess the
viability of achieving enhanced sensitivities for the detection of
gravitational waves with frequencies between $10^{-4}$ and $1$ Hz.

The aim is the design of novel gravitational antennae in
interplanetary orbit. The implementation of these ideas would be
complimentary to existing programmes of gravitational wave research
but exploiting a current niche in the frequency spectrum.

The dynamics of slender structures, several km in length, are
ideally suited to analysis by the simple theory of Cosserat
rods. Such a description offers a clean conceptual separation of
the vibrations induced by bending, shear, twist and extension and
the coupling between eigen-modes due to tidal accelerations can be
reliably estimated in terms of the constitutive properties of the
structure. The detection of gravitational waves in the 1 Hz region
would provide vital information about stochastic backgrounds in the
early Universe and the relevance of super-massive black holes to
the processes that lead to processes in the centre of galaxies.

{\bf arXiv.org}:
\href{http://arxiv.org/abs/gr-qc/0112004}{gr-qc/0112004}

\item {\bf Valiente Kroon, Juan Antonio} \enl
(jav@aei-potsdam.mpg.de, Albert--Einstein--Institut, Golm, Germany)

{\bf Early radiative properties of the developments of time
symmetric, conformally flat initial data} \enl
Using a representation of spatial infinity based in the properties
of conformal geodesics, the first terms of an expansion for the
Bondi mass for the development of time symmetric, conformally flat
initial data are calculated. As it is to be expected, the Bondi
mass agrees with the ADM at the sets where null infinity
``touches'' spatial infinity. The second term in the expansion is
proportional to the sum of the squared norms of the Newman--Penrose
constants of the spacetime.  In base of this result it is argued
that these constants provide a measure of the incoming radiation
contained in the spacetime. This is illustrated by means of the
Misner and Brill--Lindquist data sets.


\item {\bf Vera, Ra\"{u}l} \enl
(r.vera@qmul.ac.uk, School of Mathematical Sciences, Queen Mary,
University of London)

{\bf Generalisation of the Einstein--Straus/Oppenheimer--Snyder
models to anisotropic settings} \enl
We study the possibility of generalising the Einstein--Straus model
to anisotropic settings, by considering the matching of locally
cylindrically symmetric static regions to the set of $G_4$ on $S_3$
locally rotationally symmetric (LRS) spacetimes.  We show that such
matchings preserving the symmetry are only possible for a
restricted subset of the LRS models in which there is no evolution
in one spacelike direction.  These results are applied to spatially
homogeneous (Bianchi) exteriors where the static part represents a
finite bounded interior region without holes.  We find that it is
impossible to embed finite static strings or other locally
cylindrically symmetric static objects (such as bottle or
coin-shaped objects) in reasonable Bianchi cosmological models,
irrespective of the matter content.  Furthermore, we find that if
the exterior spacetime is assumed to have a perfect fluid source
satisfying the dominant energy condition, then only a very
particular family of LRS stiff fluid solutions are compatible with
this model.

Finally, given the interior/exterior duality in the matching
procedure, our results have the interesting consequence that the
Oppenheimer--Snyder model of collapse cannot be generalised to such
anisotropic cases.

{\bf arXiv.org}:
\href{http://arxiv.org/abs/gr-qc/0205011}{gr-qc/0205011}

\item {\bf Villalba, Victor} \enl
(villalba@th.physik.uni-frankfurt.de, Institut f\"{u}r
Theoretische Physik, Johann--Wolfgang--Goethe Universit\"{a}t,
Frankfurt am Main, Germany)

{\bf Creation of scalar and Dirac particles in the presence of
electromagnetic fields in cosmological backgrounds} \enl
We compute the density of scalar and Dirac particles created by a
cosmological background in the presence of homogeneous
electromagnetic fields. In order to compute the rate of particles
created we apply a quasi-classical approach.  The idea behind the
method is the following: First, we solve the relativistic
Hamilton--Jacobi equation and, looking at its solutions, we
identify positive and negative frequency modes.  Second, after
separating variables, we solve the Klein--Gordon and Dirac
equations and, after comparing with the results obtained for the
quasi-classical limit, we identify the positive and negative
frequency states and compute the Bogoliubov coefficients. We
discuss the influence of electromagnetic fields on the particle
creation process in some homogeneous cosmological models.

{\bf arXiv.org}:
\href{http://arxiv.org/abs/gr-qc/0112006}{gr-qc/0112006}

\item {\bf Waters, Thomas} \enl
(thomas.waters2@mail.dcu.ie, Dublin City University, Ireland)

{\bf Stability of Cauchy horizon in Vaidya space-time} \enl
We examine scalar radiation impinging on the Cauchy horizon in the
Vaidya space-time, and discover that an observer crossing the
horizon measures a finite flux.  We also examine the stability of
the horizon with respect to linear gravitational perturbations.


\item {\bf Watts, Anna} \enl
(alw1@maths.soton.ac.uk, Faculty of Mathematical Studies,
University of Southampton)

{\bf Stability of differentially rotating neutron stars} \enl
Previous studies have shown the oscillations of neutron stars to be
a promising source of gravitational waves.  The majority of these
studies have considered uniformly rotating stars.  Neutron stars
are however likely to be born rotating differentially.
Differential rotation complicates the stellar model as it
introduces singularities into the dynamical equations that give
rise to corotation points and a continuous spectrum.  We present
results from a simple model that highlights the key features of the
oscillations of a differentially rotating star, including possible
new instabilities.


\item {\bf Weeks, Richard} \enl
(rhw101@york.ac.uk, Department of Mathematics, University of York)

{\bf The physical graviton two-point function in de Sitter
spacetime} \enl The graviton two-point functions in de Sitter
spacetime have been calculated in various gauges.  Most of them
grow as the distance between the two points becomes large. (The
growth is usually logarithmic.)  Although it has been shown that
such growth does not translate into gauge-invariant correlation
functions which increase with distance, it will still be
interesting to find a graviton two-point function without this
behaviour.  In this work we calculate the physical graviton
two-point function in the coordinate system with $S^3$ spatial
sections which covers the whole spacetime.  This two-point function
appears not to grow as a function of the two-point distance.


\item {\bf Williams, Rhiannon} \enl
(R.L.Williams@maths.soton.ac.uk, Faculty of Mathematical Studies,
University of Southampton)

{\bf A characteristic approach to perturbed Kerr black holes} \enl
In preliminary work, we have developed a numerical code for
evolving the Regge--Wheeler equation governing black hole
perturbations in Schwarzschild spacetime.  Boundary problems are
avoided in an evolution over the entire exterior spacetime by
matching ingoing and outgoing compactified null hypersurfaces.
Following this, we have used the characteristic approach to evolve
the 3D wave equation in axisymmetric flat space.  This work is now
extended to perturbations in Kerr spacetime by evolving the
Teukolsky equation on compactified null hypersurfaces.


\end{enumerate}

\end{document}